\documentclass[11pt,twoside]{article}

\usepackage{asp2014}

\aspSuppressVolSlug
\resetcounters

\begin{document}

\title{MPDAF - A Python package for the analysis of VLT/MUSE data}
\author{L. Piqueras,$^1$ S. Conseil,$^1$ M. Shepherd$^1$, R. Bacon$^1$, F. Leclercq$^1$ and J. Richard$^1$
\affil{$^1$Univ Lyon, Univ Lyon1, Ens de Lyon, CNRS, Centre de Recherche Astrophysique de Lyon UMR5574, F-69230, Saint-Genis-Laval, France; \email{laure.piqueras@univ-lyon1.fr}}}

\paperauthor{L. Piqueras}{laure.piqueras@univ-lyon1.fr}{ORCID_Or_Blank}{Univ Lyon, Univ Lyon1, Ens de Lyon, CNRS}{Centre de Recherche Astrophysique de Lyon UMR5574}{Saint-Genis-Laval}{}{F-69230}{France}
\paperauthor{S. Conseil}{simon.conseil@univ-lyon1.fr}{ORCID_Or_Blank}{Univ Lyon, Univ Lyon1, Ens de Lyon, CNRS}{Centre de Recherche Astrophysique de Lyon UMR5574}{Saint-Genis-Laval}{}{F-69230}{France}
\paperauthor{M. Shepherd}{martin.shepherd@univ-lyon1.fr}{ORCID_Or_Blank}{Univ Lyon, Univ Lyon1, Ens de Lyon, CNRS}{Centre de Recherche Astrophysique de Lyon UMR5574}{Saint-Genis-Laval}{}{F-69230}{France}
\paperauthor{R. Bacon}{roland.bacon@univ-lyon1.fr}{ORCID_Or_Blank}{Univ Lyon, Univ Lyon1, Ens de Lyon, CNRS}{Centre de Recherche Astrophysique de Lyon UMR5574}{Saint-Genis-Laval}{}{F-69230}{France}
\paperauthor{F. Leclercq}{floriane.leclercq@univ-lyon1.fr}{ORCID_Or_Blank}{Univ Lyon, Univ Lyon1, Ens de Lyon, CNRS}{Centre de Recherche Astrophysique de Lyon UMR5574}{Saint-Genis-Laval}{}{F-69230}{France}
\paperauthor{J. Richard}{johan.richard@univ-lyon1.fr}{ORCID_Or_Blank}{Univ Lyon, Univ Lyon1, Ens de Lyon, CNRS}{Centre de Recherche Astrophysique de Lyon UMR5574}{Saint-Genis-Laval}{}{F-69230}{France}

\begin{abstract}
MUSE (Multi Unit Spectroscopic Explorer) is an integral-field spectrograph mounted on the Very Large Telescope (VLT) in Chile and made available to the European community since October 2014.
The Centre de Recherche Astrophysique de Lyon has developed a dedicated software to help MUSE users analyze the reduced data.
In this paper we introduce MPDAF, the MUSE Python Data Analysis Framework, based on several well-known Python libraries (Numpy, Scipy, Matplotlib, Astropy) which offers new tools to manipulate MUSE-specific data.
We present different examples showing how this Python package may be useful for MUSE data analysis.
\end{abstract}

\section{Introduction}

The MUSE instrument [\citet{2010SPIE.7735E..08B}] is a panoramic Integral Field Spectrograph installed on ESO's VLT UT4 $8m$ telescope since the beginning of 2014.
The fore-optics of MUSE feeds 24 identical spectrographic modules (IFU) that together cover a 1 squared arcmin field of view.
Each IFU is composed of an original advanced image slicer with a combination of 48 mirrors and 48 mini-lenses arrays, a high-throughput
spectrograph and a $4k \times 4k$ CCD.
The instrument samples almost the full spectral octave ($465-930 nm$) with $R = 2000...4000$.
Spatially, the instrument samples the sky with 0.2 arcseconds spatial pixels in the currently offered Wide Field Mode (WFM).

It was decided long before the first light of MUSE to develop a tool for the analysis of the future data.
We chose the Python langage, one of the most popular language in astronomy that proposes an important collection of libraries
(Numpy, Scipy, Matplotlib, Astropy [\citet{2013A&A...558A..33A}] ...).
MPDAF is compatible with Python 2.7 or later (including 3.x version). It runs on Linux and Mac OS X systems.

At first, MPDAF was used  for the assembly, integration tests and global tests of the instrument. Two modules were developed for that:
\texttt{mpdaf.drs} to read and manipulate the calibration products and the intermediate files of the MUSE pipeline, and \texttt{mpdaf.obj} to read and analyze the datacubes.
With exposures taken on sky, the \texttt{mpdaf.obj} module has improved considerably and there was also a need to develop tools for the source detection.
This third module \texttt{mpdaf.sdetect} is still actively developed.

In the beginning, MPDAF was only available for the MUSE consortium but recently it was released as an open-source software.
MPDAF is available through a web interface \footnote{MPDAF web interface \url{https://git-cral.univ-lyon1.fr/MUSE/mpdaf}} for software distribution (git repository) and bug/problem reporting.
It can be installed with standard tools like \texttt{pip} and its user documentation\footnote{MPDAF documentation \url{http://mpdaf.readthedocs.io/}} gives some examples of uses.

In the three next sections, we will give examples for each MPDAF sub-package.

\section{Analysing a data cube with \texttt{mpdaf.obj}}

MUSE Raw data are selected, associated and inserted through a complex reduction mechanism [\citet{2012SPIE.8451E..0BW}] which produces 3-dimensional datacubes.
Every datacube consists of 90,000 spectra each covering $465-930 nm$ with $1.25$\r{A}  step, and fully sampling
a contiguous $1'\times 1'$ field of view with $0.2'' \times 0.2''$ apertures (spaxels).
The data format follows the FITS standard. Each datacube has a 3D image with two spatial and one spectral coordinates in its first extension, the second extension contains the error information.

MPDAF provides a way to load a MUSE cube into a Python object handling the world coordinates, the variance and the bad pixels information.
Specifically, a \texttt{mpdaf.obj.Cube} object contains a 3D array of pixel values, a 3D array of variances, two attributes that describe the spectral axis and the spatial axes, and a mask array for indicating bad pixels.
The fluxes and their variances are viewed as Numpy masked arrays, so virtually all Numpy and Scipy functions can be applied to them.
But also and overall, a lot of operations can be performed directly on \texttt{Cube} propagating the world coordinates, the associated variance and the mask.
It is then relatively easy to extract smaller cubes or narrow-band images from a cube, spectra from an aperture, and perform common operations like masking, interpolating, re-sampling, smoothing, profile fitting...
The world coordinates, the associated variance and the mask are propagated into the extracted cube, image, or spectra.

For example, the following items show how MPDAF is used to caracterize a Ly$\alpha$ source in the MUSE acquisition of the Hubble Deep Field South (HDFS).
Opening the FITS as a \texttt{Cube} object, we select a sub-cube of $10''$ around the object located ar $\alpha=338.2168$ and $\delta=-60.561$:
\small\begin{verbatim}
from mpdaf.obj import Cube
cubeHDFS = Cube('DATACUBE-HDFS-v1.34.fits')
center = (-60.56183, 338.2168)
subCube = cubeHDFS.subcube(center, 10)
\end{verbatim}
\normalsize
We compute the integrated spectra within a $1''$ radius aperture, perform a fit of the line in order to measure the peak value, the FWHM ... and plot the result (Figure \ref{fig1}):
\small\begin{verbatim}
sp = subCube.aperture(center, 1)
sp.plot(title='integrated spectrum within 1" radius aperture')
sp.plot(lmin=5180, lmax=5260, title='zoom over Lya line')
lfit, rfit = sp.gauss_asymfit(lmin=5180, lmax=5260, plot=True)
\end{verbatim}
\normalsize
We reconstruct the white-light image and the Ly$\alpha$ narrow-band image by summing spatial pixels of the cube over the entire wavelength axis or over $11$\r{A}.
We then mask some data to easily compute the integrated flux and its error within a $3''$ radius aperture (Figure \ref{fig1}):
\small\begin{verbatim}
white = subCube.sum(axis=0)
nb = subCube.get_image(wave=(lfit.lpeak-5.5, lfit.lpeak+5.5))
nb.mask_region(center, 3, inside=False)
\end{verbatim}
\normalsize
\articlefigure{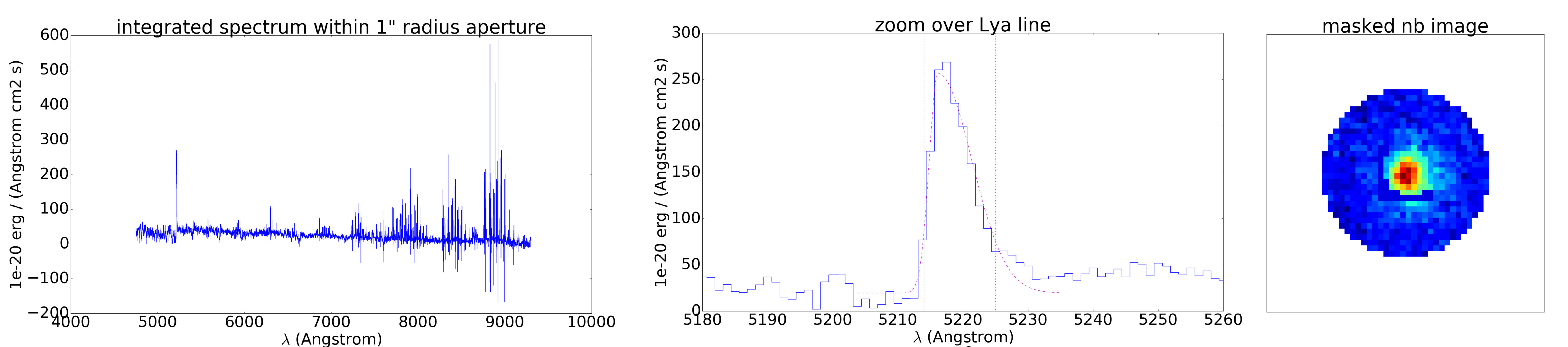}{fig1}{\emph{Left:} integrated spectrum. \emph{center:} fit of the Ly$\alpha$ line. \emph{Right:} circular mask on the narrow-band image.}

The study led by [Leclercq and al (in preparation)] makes a good use of the MPDAF package.
It consists of the analysis of the circum-galactic medium of individual high redshift galaxies using the Ly$\alpha$ line detected with MUSE.

\section{\texttt{mpdaf.sdetect}: offering tools to detect and manage sources}

A lot of tools are currently been developed to detect sources in MUSE data cubes.
The \texttt{mpdaf.sdetect} module should provides different methods of detections.
For now, MPDAF provides MUSELET, a SExtractor-based tool to detect emission lines in a data cube, which has been used for several MUSE studies within the consortium [\citet{2016arXiv160902920D}, \citet{2016ApJ...820..121B}, \citet{2016A&A...590A..14B}]

We are working on a format to gather all the information on a source in one FITS file in order to easily exchange information about detected sources.
Generic information are stored in the primary header of the file, FITS image
extension are used to store spectra, small images and sub data cubes and FITS
binary table extensions are used to store the information relative to line profiles,
magnitudes and redshift values. The \texttt{Source} class implements input/output for this source FITS file.
Returning to the example of the Ly$\alpha$ line detected in the MUSE HDFS cube, its FITS file contains the sub-cube, the full spectrum (and the zoom on the emission line),
the values of the flux, the FWHM, the observed wavelength, the white light image, the continuum image, the narrow-band image, the mask used to hide the others sources, the corresponding HST image ...
All images are saved in a dictionary of \texttt{mpdaf.obj.Image} objects. Figure \ref{fig_ima} shows some images of this source.

\articlefigure{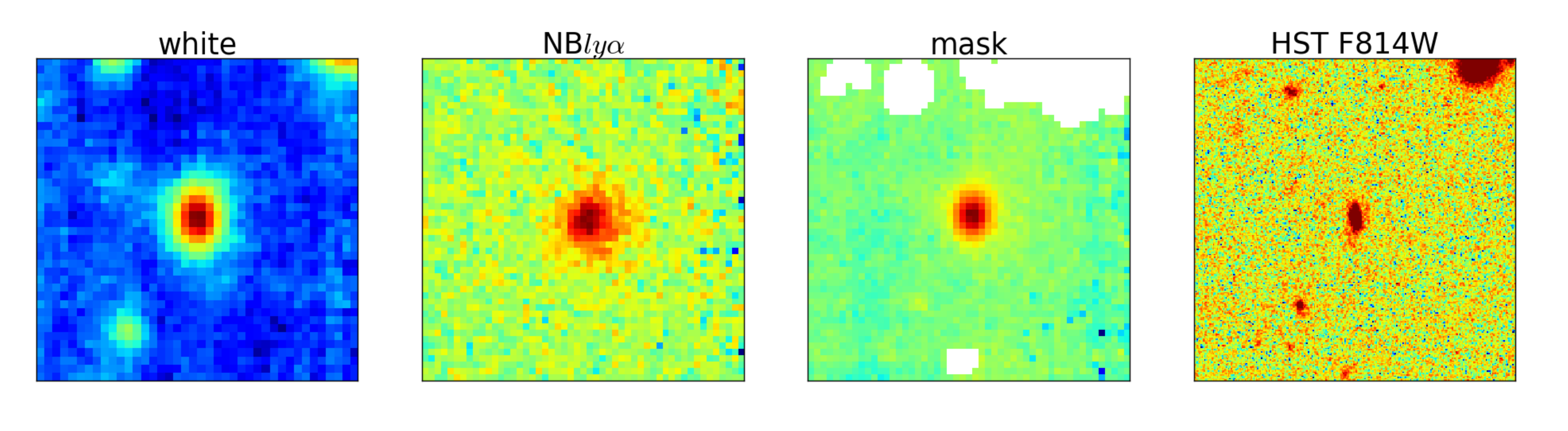}{fig_ima}{Different images stored in the source file.}

\section{Working on MUSE pixel table object with \texttt{mpdaf.drs}}

In the reduction approach of the MUSE pipeline, data need to be kept un-resampled until the very last step.
The pixel tables used for this purpose can be saved at each intermediate reduction step.
It is a tabular format where a row corresponds to each CCD pixel and the value of each pixel is stored together with its coordinates (position and wavelength).
Its size is up to 9GB (a pixtable containing a MUSE field for a single exposure has 323885723 rows).

The \texttt{mpdaf.drs.PixTable} class is used to handle the MUSE pixel tables created by the data reduction system.
Several methods exist to extract a subset of a pixtable using spatial, spectral or origin (IFU, slice, CCD) criteria.
These methods can be used to investigate how the MUSE field is cut by the instrument and how it is mapped on the detector.
By so doing, [\citet{2015A&A...575A..75B}] compute the median value for all slices and correct the pixtable to remove residual offsets between slices before reconstructing the data cube.
For details, refer to [\citet{P1.8_adassxxvi}].

\section{Conclusion}

MPDAF provides tools to work with MUSE-specific data and with more general data like spectra, images and data cubes.
Although its main use is to work with MUSE data, it is also possible to use it other data, for example HST images.

It has been developed and used in the MUSE Consortium for several years, and is now available freely for the community.
We encourage users to report bugs, to provide comments and to use or adapt this code as they see fit.

\acknowledgements R. Bacon acknowledges support from the ERC advanced grant 339659-MUSICOS.

\bibliography{P6-21}  

\end{document}